# Radiation tolerance: Nano triumphs bulk

Parswajit Kalita[1,*], Santanu Ghosh[1,*], Gaëlle Gutierrez[2], Parasmani Rajput[3], Vinita Grover[4], Gaël Sattonnay[5], Devesh K. Avasthi[6,7*]

[1]*Dept. of Physics, Indian Institute of Technology Delhi, New Delhi – 110016, India*
[2]*CEA Saclay, DEN, SRMP, Labo JANNUS, 91191 Gif-sur-Yvette, France*
[3]*Atomic & Molecular Physics Division, Bhabha Atomic Research Centre, Mumbai – 400085, India*
[4]*Chemistry Division, Bhabha Atomic Research Centre, Mumbai – 400085, India*
[5]*LAL, Université Paris-Sud, Bât 200 F-91405, Orsay, France*
[6]*Amity Centre for Accelerator based Applied & Fundamental Research, Amity University, Noida – 201313, India*
[7]*Amity Centre for Advance Research & Innovation, Amity University, Noida – 201313, India*

*Materials are subjected to energetic particles in a number of radiation environments, and are hence prone to undesirable (radiation) damage. We report here the superiority of the nano-crystalline phase over bulk for radiation tolerance under simultaneous irradiation with high energy (electronic energy loss ($S_e$) dominant) and low energy (nuclear energy loss ($S_n$) dominant) particles. Nano-crystalline yttria stabilized zirconia is found to exhibit lesser radiation damage (viz. degradation in crystallinity), when compared to its bulk-like counterpart, against simultaneous irradiation with high energy 27 MeV Fe and low energy 900 keV I ions. This is interpreted within the framework of the 'thermal-spike' model after considering (i) the fact that there is essentially no spatial and time overlap between the damage events of the two 'simultaneous' ion beams, and (ii) the influence of grain size on the radiation damage against separate $S_n$ and $S_e$. The present work besides being of keen interest for fundamental understanding of ion-material interactions, also paves the way for the potential application of nano-crystalline materials in the nuclear industry where such simultaneous irradiations are encountered.*

*Correspondence email(s): phz148112@physics.iitd.ac.in (PK), santanu1@physics.iitd.ac.in (SG), dkavasthi@amity.edu (DKA)

## 1. Introduction

Exposure of materials to energetic particles / ions, i.e. irradiation, often results in the creation of defects and subsequent micro-structural changes in the material, eventually leading to a degradation of its properties (i.e. radiation damage). Such exposure of materials to energetic particles is thus a detrimental process. This is a crucial issue in several vital fields of science and technology (e.g. nuclear, electronic and space industries[1-3]) where the materials are subjected to severe irradiation with low energy (slowing down primarily by ballistic/nuclear collisions ($S_n$)[4]) and/or high energy (predominantly slowing down by electronic excitations ($S_e$)[4]) particles over time. For these applications, understanding the behaviour of the materials under the corresponding in-service conditions, and thus designing ways to make them more radiation tolerant, is therefore crucial. A common way to simulate the effects of such irradiations, within a limited time, is to use external ion beams obtained from (ion) accelerators.

In the recent past, nano-scale materials design, i.e. downsizing of materials to nano-dimension, has received significant attention in the context of mitigating the radiation damage in materials. It has been observed that the nano-crystalline (NC) state exhibits enhanced radiation tolerance, against low energy ions ($S_n$ dominant), when compared to its bulk counterparts[5-13]. We have, however, very recently shown that the better radiation tolerance of NC materials is not always true and designing nano-scale materials in order to lower the irradiation induced damage is not advantageous against high energy ions ($S_e$ dominant)[14]. In other words, our results[14] show that the situation is very different under high energy irradiations and the NC phase is more damaged as compared to its bulk-like counterpart. A similar behaviour has been observed in the case of Ceria irradiated with high energy ions as well[15]. Note here that the mechanisms of $S_n$ and $S_e$ are fundamentally very different. This opposite dependence of the radiation tolerance on the grain size, based on the energy loss mechanism (i.e. $S_n$ or $S_e$) of the incident particles, thus raises an intriguing question, viz. what will be the effect of grain size on the radiation tolerance against simultaneous $S_n$ and $S_e$? Will the effect be similar to that against low energy irradiations and the NC state be more radiation tolerant; or will the effect be more inclined towards the high energy irradiation results and the NC phase be more damaged? In either case, why? Or will the effect be something totally different (and unexpected)?

Note that understanding the grain size mediated radiation response under concomitant $S_n$ and $S_e$, apart from being of fundamental interest, is also mandatory from an application point of view since materials used in nuclear reactors (e.g. IMFs[16]) and/or waste matrices are actually exposed to simultaneous irradiation with low (alpha recoils) and high (fission fragments) energy particles. Here, the importance of nuclear energy in fulfilling our (energy) requirements, particularly in the light of rapidly depleting fossil fuel reserves and climate change, is a point worth considering.

Despite the decades of research devoted to understanding the radiation damage in materials, studies concentrating on the effects of simultaneous $S_n$ and $S_e$ irradiations are very scarce[1, 2, 17, 18]; i.e. the effects of single and sequential irradiations with low and/or high-energy ions are well documented (see e.g. Refs.[1, 4, 19] and references therein), but the same is not the case with simultaneous particles irradiations. In a recent pioneering work, Thome et al.[1] had reported the effects of simultaneous $S_n$ and $S_e$ irradiation on various classes of materials. More recently, simultaneous irradiation studies on silica and $UO_2$ have been reported[2, 18]. The materials in the study[1] of Thome and co-workers were all in the *single-crystalline* state, and moreover the studies were, to our best understanding, aimed at examining (possible) combined/cooperative effects of $S_n$ and $S_e$. On the other hand, the silica, in the report by Mir et al.[2], was in the *amorphous* form, and furthermore the aim of this particular study was, in our best interpretation, to understand what makes single beam and/or sequential irradiation scenarios different from a simultaneous irradiation scenario, and if there exists any space-time correlation effects during simultaneous irradiation that modify the damage evolution process. Similarly, the study in Ref.[18] dealt with the effect of coupled electronic and nuclear energy deposition on strain and stress levels on $UO_2$ *polycrystals*. It is thus quite apparent that (i) the question (regarding the influence of the grain size on the radiation tolerance against simultaneous $S_n$ and $S_e$) that we are trying to address here is fundamentally very different (from those in these studies[1, 2, 17, 18]), and hence (ii) such a problem has not at all been investigated earlier.

The investigation of this intriguing question constitutes the work reported in this letter. Yttria stabilized zirconia (YSZ) is chosen in the current work because of its importance in the nuclear industry[16, 20-22] and because a good amount of research has already been done (single ion beam irradiations) thus making available vital results for comparison. Note, again, that in the context of nuclear materials, simultaneous irradiations with high energy and low energy particles (and not single / separate low or high energy irradiations) correspond to the in-service conditions of actual relevance.

## 2. Experimental details

10 mol% YSZ powder was prepared by gel combustion method (see Ref.[23] for details) and then compacted into pellets of diameter ~ 8 mm. The pellets were subsequently heated at 600ºC for 6 hours and 1300ºC for 84 hours with the aim of obtaining different microstructures / crystallite sizes. The pellets heated at 600ºC and 1300ºC are hereafter referred to as S600 and S1300 respectively.

The S600 and S1300 samples were then *simultaneously* irradiated with 27 MeV Fe ions ($S_e$ dominant) and 900 keV I ions ($S_n$ dominant) at the JANNUS-Saclay facility[24, 25]. The fluence of both the ion species is $10^{15}$ ions/cm$^2$; the irradiations are performed at room temperature with the ion fluxes limited to ~ $10^{11}$ ions/cm$^2$/sec. The S600 and S1300 samples were also irradiated with *only* the 900 keV I ions keeping all conditions same as in the simultaneous irradiations. The electronic energy loss ($S_e$), nuclear energy loss ($S_n$) and the projected range of the 27 MeV Fe ions are estimated to be ~ 12 kev/nm, ~ 0.07 keV/nm and ~ 4.3 ± 0.4 µm respectively by SRIM[26] simulation code; while the corresponding values for the 900 keV I ions are ~ 1.2 keV/nm, ~ 3.1 keV/nm and ~ 185 ± 55 nm respectively.

GIXRD and Raman spectroscopy measurements, of pristine and irradiated pellets, were performed for structural investigations, including the changes / damage induced by the irradiations. The GIXRD patterns were recorded with Cu K$_α$ radiation using a Philips X'Pert Pro diffractometer. The incidence angle was fixed at 0.5º; the probed depth is ~ 140 nm at this angle of incidence. The Raman spectra are recorded using a confocal Renishaw InVia Raman microscope with a laser excitation wavelength of 514 nm and spot size of ~ 1 µm$^2$ on the sample surface. The phase of both S600 and S1300 was verified to be the cubic phase by GIXRD and Raman spectroscopy; the average crystallite (grain) size of pristine S600 and S1300 is determined to be ~ 26 nm and ~ 80 nm from the XRD peak broadening (see Ref.[14] for details) and TEM (Figure 1). SEM imaging (Figure 1) of the pristine samples revealed S1300 to be highly dense and having well-defined particles of size ~ 4.5 ± 2.4 µm, while S600 was found to be less dense and having a much smaller particle size of ~ 38 ± 9 nm. The S1300 sample can thus be considered as bulk. EXAFS was used to probe the local structural order / environment in the samples, i.e. to investigate the structural changes induced upon irradiation at the atomic scale. The EXAFS measurements were performed at the zirconium (Zr) K-edge in fluorescence mode at the Scanning EXAFS beamline (BL-9) at RRCAT, India. The detector was set at the smallest possible glancing angle (with respect to the sample surface, less than 5º) to ensure that only the fluorescence photons arising from the near surface region are collected. The energy range was calibrated using Zr metal foils. The FEFF 6.0 code[27] has been used to analyse the EXAFS data. The code facilitates background removal and Fourier transform for deriving the $\chi(R)$ versus $R$ spectra from the absorption data (using ATHENA software). This is followed by generation of a theoretical XAFS spectra from an assumed crystallographic structure and lastly the fitting of the experimental data with the theoretical spectra via ARTEMIS software[28].

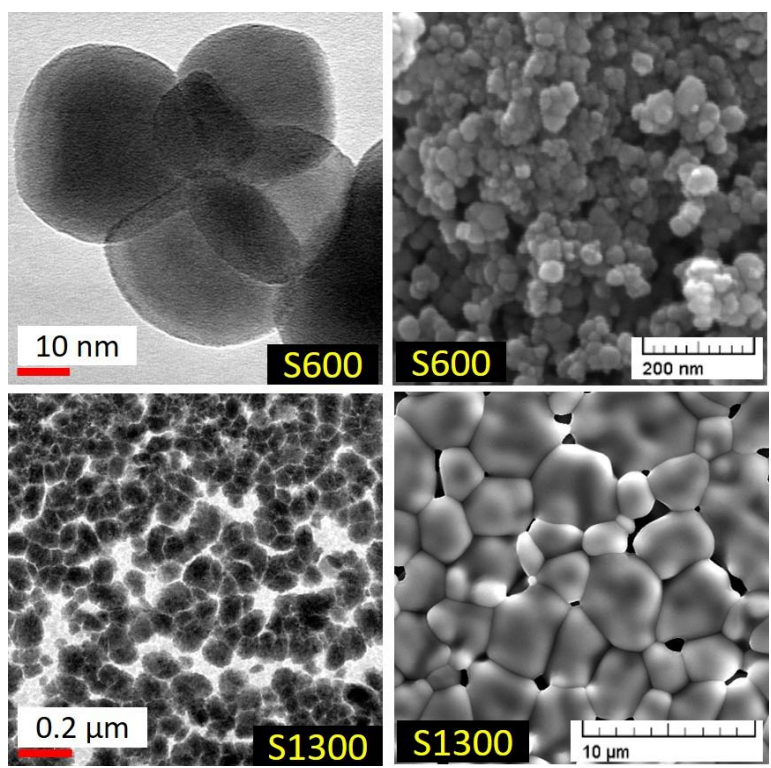

**Figure 1:** TEM (left column) and SEM (right column) images of pristine S600 and S1300. Adapted from Ref.[14].

## 3. Results
### 3.1 X-ray diffraction

The evolution of the GIXRD patterns of the S600 and S1300 samples irradiated simultaneously with the 27 MeV Fe and 900 keV I ions are shown in Figure 2. The evolution of the GIXRD patterns upon only the 900 keV I irradiation is also shown here. The XRD patterns reveal that, irrespective of the crystallite size and/or type of irradiation (single beam or simultaneous), the XRD peak broadening has increased upon irradiation. This indicates that the irradiations have resulted in the degradation of the crystallinity (i.e. damage). Now in order to evaluate the influence of the grain size on the irradiation induced damage (or conversely the radiation tolerance) against simultaneous $S_n$ and $S_e$, it is first necessary to have a quantitative estimate of the radiation damage in the various cases. Since the XRD peak broadening is a measure of the (degradation in) crystallinity, the irradiation induced damage is quantified by the relative change in the FWHM upon irradiation. The irradiation damage is accordingly estimated using the equation

$$damage = \frac{FWHM_{(111)\_irradiated} - FWHM_{(111)\_pristine}}{FWHM_{(111)\_pristine}} \qquad \ldots\ldots (1)$$

where $FWHM_{(111)\_pristine}$ & $FWHM_{(111)\_irradiated}$ are the FWHM of the (111) diffraction peak for the pristine and irradiated samples respectively. The FWHM of the (111) peak for the pristine and irradiated S600 and S1300 samples, and the corresponding damage are summarized in Table 1. From these values, it is, firstly, apparent that the NC S600 sample is significantly less damaged, under the 900 keV I ($S_n$) irradiations, when compared with the bulk-like S1300 sample. This is in excellent agreement with previous literature (see e.g. Refs.[5, 10]) on the effect of grain size on the radiation damage against low energy ($S_n$) irradiations. Secondly, irrespective of the crystallite size, the damage after the simultaneous irradiations is greater than the damage after the *single* 900 keV I irradiations. This indicates that no cooperative $S_n/S_e$ effects, that induces a healing of the $S_n$ induced damage (like in SiC[1, 17] and MgO[1]) are observed during the simultaneous irradiations in these cubic zirconia samples. This result is again in good agreement with previous literature[1]. Apart from these somewhat expected findings, the important (and very interesting) observation is that the NC S600 sample is significantly less damaged under the simultaneous 900 keV I ($S_n$) and 27 MeV Fe ($S_e$) irradiations. In other words, the XRD results indicate that the NC sample is more radiation tolerant than its bulk (-like) counterpart against the simultaneous $S_n$ and $S_e$ irradiations. This result will be addressed later, in detail, in connection with the general nature of simultaneous irradiations coupled with the role of grain size on the radiation damage against separate $S_n$ and $S_e$.

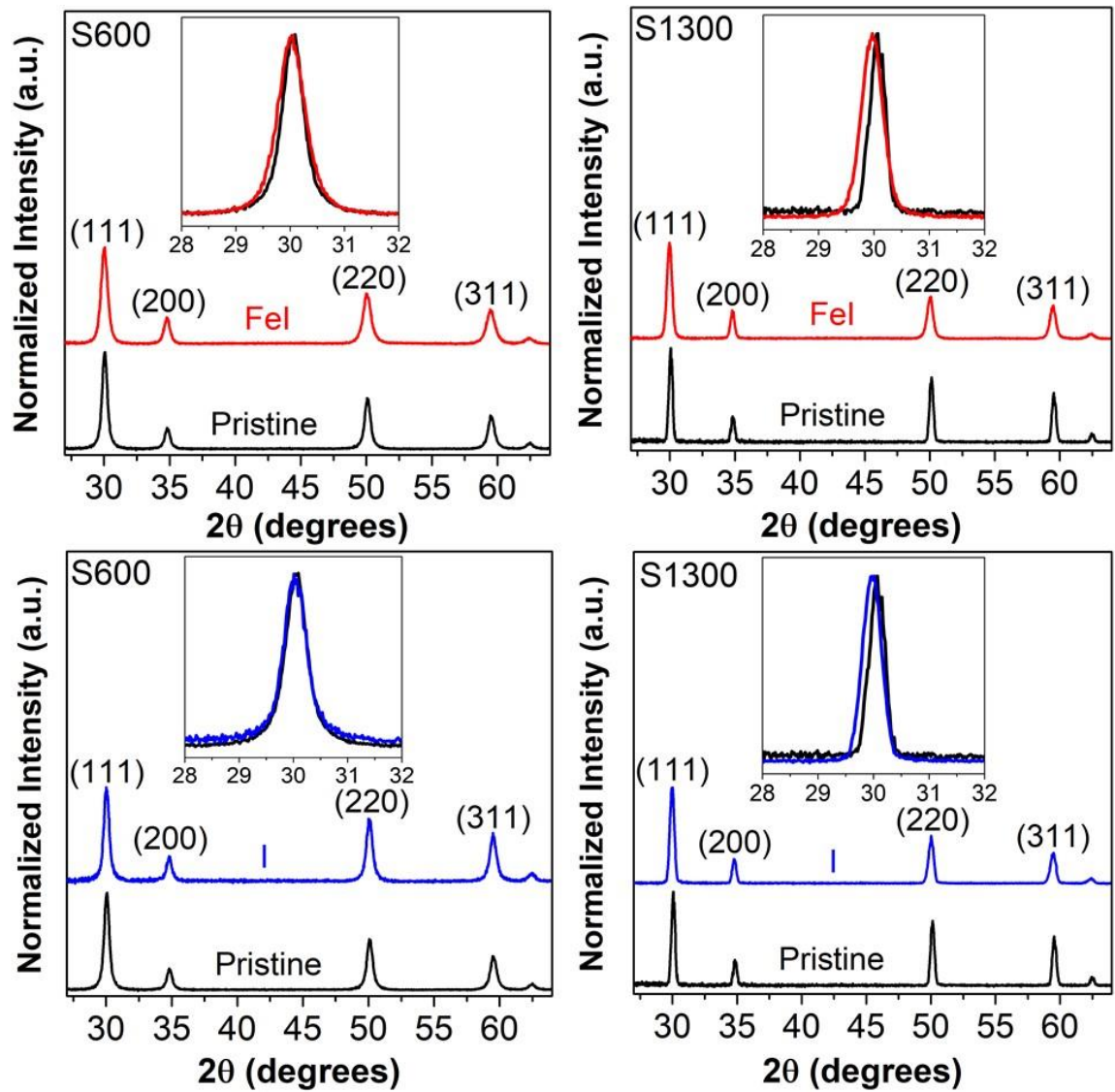

**Figure 2:** GIXRD patterns of pristine and irradiated S600, S1300. The top row corresponds to the simultaneous (FeI) irradiation, bottom row corresponds to the *single* 900 keV I irradiations. Magnified view of the (111) peak for all samples is shown as inset.

|  | $FWHM_{(111)}$ (degree) | Damage (XRD) | $FWHM_{F2g}$ (cm$^{-1}$) | Damage (Raman) |
|---|---|---|---|---|
| S600 pristine | 0.52 ± 0.02 |  | 96 ± 1 |  |
| S600 I | 0.56 ± 0.02 | 7.7 % | 100 ± 1 | 4.2 % |
| S600 FeI | 0.63 ± 0.02 | 21.1 % | 106 ± 1 | 10.4 % |
| S1300 pristine | 0.31 ± 0.02 |  | 80 ± 1 |  |
| S1300 I | 0.39 ± 0.02 | 26 % | 89 ± 1 | 11.3 % |
| S1300 FeI | 0.46 ± 0.02 | 48.4 % | 98 ± 1 | 22.5 % |

**Table 1:** FWHM of (111) XRD peak, FWHM of $F_{2g}$ Raman band and the damage as calculated from XRD and Raman spectroscopy for all samples. FeI denotes the simultaneous irradiations, I denotes the *single* 900 keV I irradiations.

## 3.2 Raman Spectroscopy

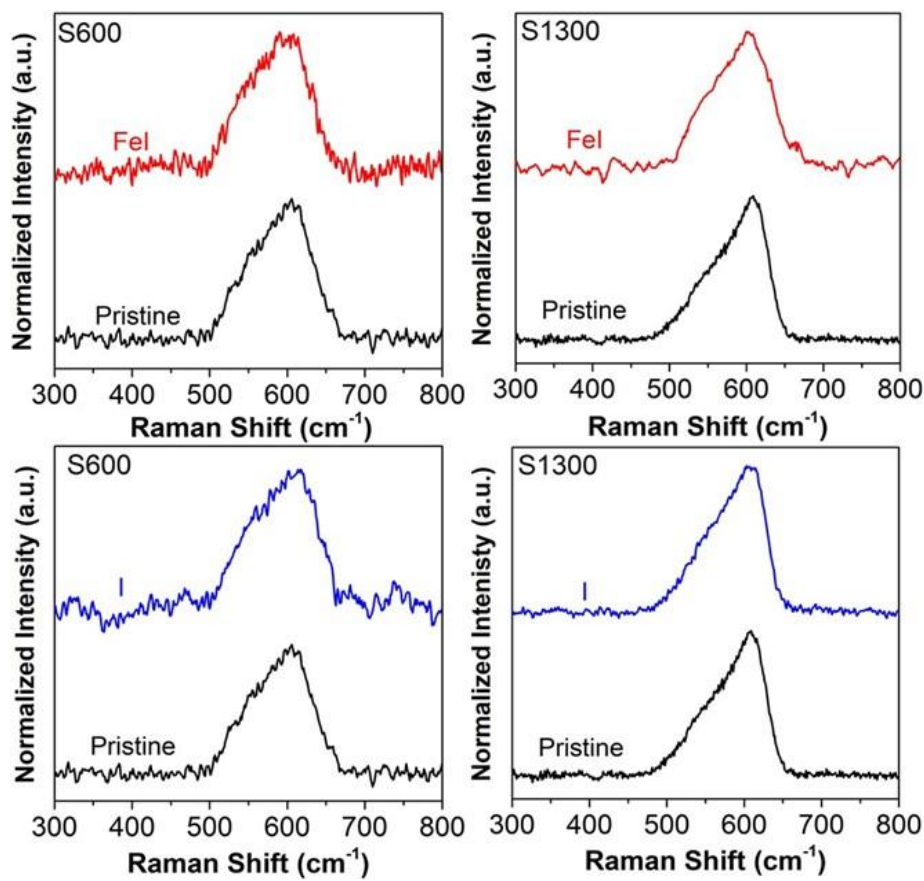

**Figure 3:** Raman spectra of pristine and irradiated S600 and S1300 showing the $F_{2g}$ band. The top and bottom rows correspond to the simultaneous (FeI) and the *single* 900 keV I irradiations respectively.

The Raman spectroscopy measurements of the pristine and irradiated samples were carried out with the aim of investigating the changes brought upon at the microscopic level upon irradiation and to substantiate the GIXRD findings. Figure 3 shows the Raman spectra of unirradiated and irradiated NC S600 and bulk-like S1300 samples. The broad asymmetric band, centred at around 608 cm$^{-1}$, seen in both these unirradiated samples is assigned to the Raman active $F_{2g}$ mode of Zr-O vibration with $O_h^5$ symmetry[29, 30]. This asymmetry is attributed to the presence of disorder in the YSZ system induced by the doping of Yttrium ($Y^{3+}$) ions into the zirconia structure[30]. The FWHM of the $F_{2g}$ band for pristine and irradiated samples are listed in Table 1. Amongst other factors, since the width of the Raman band is a measure of the crystallinity, it is apparent that irradiation has resulted in degradation of the crystallinity (peak broadening), i.e. damage, irrespective of the crystallite size and/or type of irradiation. The damage was again quantified using equation (1) with the FWHM of $F_{2g}$ band of pristine and irradiated sample, and is listed in Table 1. In agreement with the XRD results, it can be seen that: (i) the NC S600 sample is significantly less damaged than the bulk-like S1300 sample under the 900 keV I irradiations, (ii) for both S600 and S1300, the damage after the simultaneous 900 keV I ($S_n$) and 27 MeV Fe ($S_e$) irradiations is higher than that after the *single* 900 keV I irradiation, and (iii) crucially, the NC S600 sample is significantly less damaged under the simultaneous 900 keV I and 27 MeV Fe irradiations as compared to the bulk-like S1300 sample. It may be noted that the damage values as estimated from Raman spectroscopy is lower than the corresponding values estimated via GIXRD. This may be due to the reason that the information depth of the two techniques are different – the Raman penetration depth is expected to be more than the ~ 140 nm probed by GIXRD and some part of the detected Raman signal may have originated from the part of the samples untouched by the 900 keV I beam. Nevertheless, the important point is that the trend in the values, estimated via both techniques, is the same.

### 3.3 EXAFS measurements

EXAFS was used to obtain information about the changes in the local atomic structure, around a Zr atom, upon irradiation. The magnitude of the Fourier transform (FT) of the $k^2$ weighted normalized EXAFS functions (i.e. $k^2\chi(k)$), for the pristine and irradiated S600 and S1300 samples, is shown in Figure 4. The $k$-range of 3 – 7.5 Å$^{-1}$ was used for the FT. Before the FT, the EXAFS function $\chi(k)$ is obtained from the energy dependent absorption function $\chi(E)$ using the relation $k = \sqrt{\frac{2m(E-E_0)}{\hbar^2}}$, where $m$ is the mass of electron, $E_0$ is the absorption edge energy and $\hbar$ is the Planck's constant. Figure 4 also shows the best fit of the magnitude of the FT; the fitting was performed using cubic zirconia structure in the $R$-space range of 1 – 3.6 Å corresponding to the first and second nearest neighbour coordination shells. The local structural parameters, i.e. coordination numbers (CN), bond distances (R) and Debye-Waller (DW) factor ($\sigma^2$), as determined by these fittings are listed below in Table 2. In line with previous literature[31, 32], the CN of the pristine samples (both S600 & S1300) was held fixed at 8 and 12 respectively for the first and second coordination shells. On the other hand, the CN were floating in the fitting of the irradiated samples since the irradiations can result in the displacement of atoms from their regular sites. Bond distances and DW factors were kept floating in the fitting procedure for all samples.

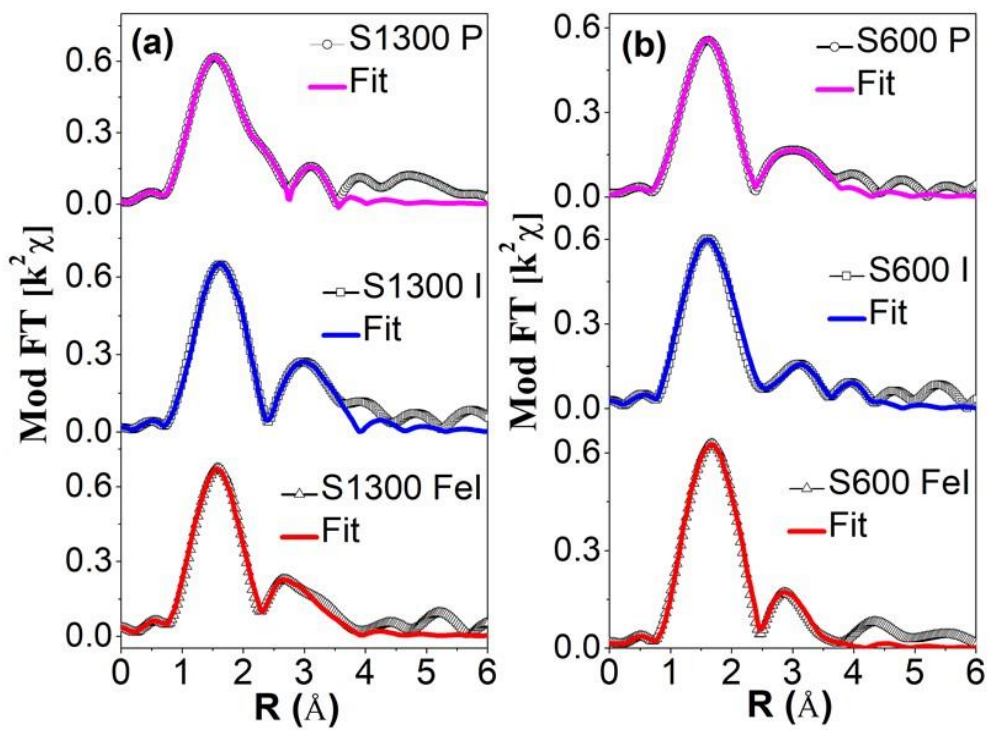

**Figure 4:** Magnitude of the Fourier transform (FT) of EXAFS functions ($k^2\chi(k)$) for pristine and irradiated S600 and S1300. S1300 P and S600P are the respective pristine samples.

|  | $CN_{Zr-O}$ | $R_{Zr-O}$ (Å) | $\sigma^2_{Zr-O}$ (Å$^2$) | $CN_{Zr-Zr}$ | $R_{Zr-Zr}$ (Å) | $\sigma^2_{Zr-Zr}$ (Å$^2$) |
|---|---|---|---|---|---|---|
| S600 P | 8 | 2.155(4) | 0.0156 (4) | 12 | 3.511(4) | 0.0167 (5) |
| S600 I | 7.3(3) | 2.221(3) | 0.0202 (4) | 10.4 (4) | 3.543(5) | 0.0218 (5) |
| S600 FeI | 6.3(4) | 2.246(4) | 0.0213 (5) | 8.5 (5) | 3.573(5) | 0.0226 (6) |
| S1300 P | 8 | 2.134 (3) | 0.0118 (3) | 12 | 3.486 (5) | 0.0127 (4) |
| S1300 I | 6.4(4) | 2.154 (4) | 0.0157 (4) | 10.7 (5) | 3.454 (4) | 0.0171 (6) |
| S1300 FeI | 5.5(3) | 2.165 (4) | 0.0179 (3) | 3.4 (5) | 3.459 (6) | 0.0194 (5) |

**Table 2:** Local structural parameters obtained from fitting of the experimental EXAFS data. The uncertainty in the last digit is indicated by the numbers in parentheses. FeI denotes the simultaneous irradiations, I denotes the single 900 keV I irradiations.

Note that ideally in bulk cubic zirconia, zirconium has 8 nearest neighbouring oxygen atoms at a mean distance of 2.204 Å followed by 12 zirconium next nearest neighbour atoms at a mean distance of 3.599 Å[33]. It is therefore apparent that the local structure of the pristine S600 and S1300 samples is not the same as bulk cubic zirconia – the bond distances are significantly shorter than the corresponding values for bulk cubic zirconia. Moreover, comparing the local structural parameters of the pristine S600 and S1300 samples, we find that the DW factor, which is indicative of the local disorder, is higher for both the coordination shells of the NC S600 sample. Upon irradiation, changes in the short-range order are evident for both S600 and S1300. These changes include a decreased CN (Zr-O and Zr-Zr) and an increased DW factor (Zr-O and Zr-Zr). Such irradiation induced changes in the local structure, i.e. decrease in CN and increase in DW factor, are well in agreement with previous literature[31, 34, 35]. The decrease in the CN indicates vacancy-like (i.e. point) defects, where some of the O and Zr atoms are displaced from their regular sites (neighbouring the *absorbing* Zr atom) by the irradiations. The increase in the DW factor upon irradiation implies an irradiation induced increase in the local disorder. The increase in the DW factor may itself be, in part, due to the creation of the point defects upon irradiation. At the same time, the increase may also be in part due to the enhancement in the thermal vibration of the O and Zr atoms upon irradiation[31]. As the current EXAFS measurements were performed only at room temperature, it is not possible to separate these two contributions to the DW factor[31]. Nonetheless, the EXAFS measurements indicate the creation of local structural damage (creation of vacancy-like / point defects and/or increase in the disorder) by the irradiations. A thorough observation of the CNs and DW factors reveal that the trend in the irradiation induced damage in the short-range (atomic scale) is the same as the trend of the irradiation damage in the long range (as indicated by GIXRD).

## 4. Discussion

The lower radiation damage in the NC S600 sample as compared to the bulk-like S1300 sample under the 900 keV I irradiations is as per expectations. Since grain boundaries (GBs) are defect sinks, the defects that are produced in the collision cascades upon the $S_n$ irradiations are trapped by them (directly (and primarily) in the collision cascades and as well as because of defect migration from nearby regions to the GBs[10, 22]) which results in their removal/reduction. This process of defect removal/reduction by the GBs is much more efficient in S600 as compared to S1300 because: (i) the volume fraction of GBs is higher in the S600 sample because of its smaller grain size, and (ii) the possibility of the irradiation induced collision cascades occurring 'near' the GBs is also higher (again because of its smaller grain size) and hence the defects can more easily interact with the GBs. Moreover, the probability of defects (that are not captured directly in the collision cascades) reaching the GBs after migrating from the nearby regions is also higher for S600. As such the $S_n$ induced defect concentration is lower in the S600 sample, as compared to S1300, and hence the radiation damage is lesser. In a detailed TEM study on the effect of low-energy 400 keV Kr ($S_n$ dominant) irradiations on different grain-sized samples of YSZ, the authors had (experimentally) shown that the defects concentration, after irradiation, is significantly lower in the NC samples as compared to the bulk sample[10]. We can expect similar behaviour in the present case given that the 900 keV I irradiations are in the low-energy ($S_n$ dominant) regime as well.

Now, in order to interpret the simultaneous irradiation results, it is first and foremost necessary to understand the (general) nature of simultaneous irradiations itself. A simultaneous irradiation scenario is, strictly speaking, simultaneous only when the second ion impact occurs at the same place and in a time period corresponding to the lifetime of the damage event from the previous (i.e. first) ion impact, such that the energy deposition by the second ion perturbs the damage event from the first ion. In other words, a simultaneous irradiation scenario is truly simultaneous only when there is both spatial and time overlap of defect formation/evolution between the two ion beams. It has however been reported that there is essentially no such space-time overlap during simultaneous irradiation[1, 2, 17] – the probability of such an overlap is actually only about ~ $10^{-11}$ (see e.g. Ref.[17]). Hence, under simultaneous irradiations, the 'effect' of the second ion would not perturb the 'effect' of the first ion (during its life-time) and would instead only, in-situ and step-by-step, *follow* it. Therefore, in this structure, simultaneous irradiation can be essentially considered to be equivalent to a series of random sequential irradiations, where the cumulative fluence of these sequential irradiations is equal to the fluence as in the simultaneous irradiation. In-fact, it has been experimentally shown (very recently) that the simultaneous irradiation scenario is indeed equivalent to multiple small sequential irradiation scenarios[2]. We have therefore interpreted our results within this well-established framework. Accordingly, the simultaneous irradiation (I & Fe) in the present case is considered as a series of irradiations with the I and Fe ions as follows

$$I\ (\Phi^I)\ \&\ Fe\ (\Phi^{Fe}) \sim I\ (\Phi^I_1) + Fe\ (\Phi^{Fe}_1) + I\ (\Phi^I_2) + Fe\ (\Phi^{Fe}_2) + \ldots + I\ (\Phi^I_n) + Fe\ (\Phi^{Fe}_n)$$

where, $\Phi^I$ and $\Phi^{Fe}$ are the fluence of the 900 keV I and 27 MeV Fe ion beams in the siumultaneous irradiations ($\Phi^I = \Phi^{Fe} = 10^{15}$ ions/cm$^2$), $\Phi^I_j$ and $\Phi^{Fe}_j$ (j = 1 to n) are the incremental fluences of I and Fe respectively that sequentially make up the simultaneous irradiation, and therefore $\Sigma\ \Phi^I_j = \Phi^I$ and $\Sigma\ \Phi^{Fe}_j = \Phi^{Fe}$.

Now, in the case of the simultaneous irradiations, the NC S600 sample will be less damaged as compared to the bulk-like S1300 sample against the 900 keV I ions (fluence $\Phi^I_1$) because of its larger fraction of GBs (that act as sinks for the defects created in the collision cascades). This is familiar from literature. The GIXRD (and Raman spectroscopy) results also show that the S600 sample is significantly less damaged than S1300 after the (*single*) 900 keV I irradiations (although these results are at the total fluence of $10^{15}$ ions/cm$^2$, the trend is expected to be the same after $\Phi^I_1$). Upon the subsequent arrival of the 27 MeV Fe ions (fluence $\Phi^{Fe}_1$), the S600 sample is expected to be more damaged than the S1300 sample because of its smaller grain size that results in a more intense thermal spike (see Ref.[14] for details). The situation is however not as straightforward. Defects in the crystalline sysytem can scatter electrons and phonons, thereby resulting in the decrease of lattice thermal conductivity ($K$) and increase in the electron-phonon coupling strength ($g$)[36]. This in turn would result in a stronger thermal spike in such a defected system as compared to a defect-free (or less defected) system. As shown by Weber et al.[37] and Liu et al.[36], the (pre-) existence of defects in a crystalline system can greatly enhance its sensitivity to electronic energy loss and thereby result in a significantly higher damaged state, upon $S_e$ irradiation, as compared to a defect-free system. In the present case, S1300 is significantly more damaged (i.e. more defects) than S600 by the I ions (fluence $\Phi^I_1$). Therefore, the intensity of the thermal spike in S1300 after the subsequent arrival of the Fe ions (fluence $\Phi^{Fe}_1$) may be comparable to (or may even be stronger than) that for S600. In other words, although the crystallite size of S600 is much smaller than S1300, the thermal spike generated in it upon Fe irradiation (fluence $\Phi^{Fe}_1$) maybe comparable to that in S1300 because of the existence of significantly more defects in the S1300 system (created earlier by $\Phi^I_1$). Note that the pre-existing defects not only influence the thermal spike but also play a crucial role in the final damage evolution[37, 38]. As such, the S1300 sample is affected by both the I ($S_n$) and Fe ($S_e$) ions, whereas the S600 sample is essentially affected only by the Fe ($S_e$) ions. In other words, the damage in S600 is effectively due to $S_e$ alone, whereas the damage in S1300 is a superimposition of the damage due to $S_n$ and damage due to $S_e$ (which itself is a consequence of the pre-existing $S_n$ damage). Therefore, the damage in S1300 is higher than in S600 after the I (fluence $\Phi^I_1$) + Fe (fluence $\Phi^{Fe}_1$) impact. This process is repeated with the subsequent series of I and Fe ions (fluence $\Phi^I_j$ + fluence $\Phi^{Fe}_j$, j = 2 to n). The net result is that S1300 is more damaged than S600 (as evident in GIXRD and Raman spectroscopy).

Thus, the higher radiation damage in the bulk-like sample is ultimately due to its larger grain size (lower density of GBs) that resulted in greater damage against $S_n$, and which in turn resulted in significant damage against $S_e$ as well. Conversely, the better radiation tolerance of the NC sample is a consequence of its very nature (i.e. nano-crystallinity) itself.

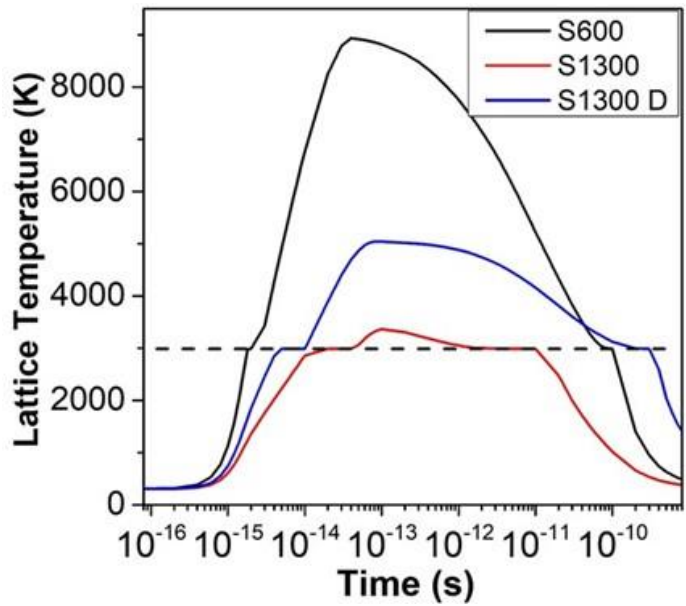

**Figure 5:** Variation in lattice temperature with time at the center of the ion track (0 nm) for pristine S600, pristine S1300 and $S_n$ defected S1300 (S1300 D). The dashed horizontal line shows the melting temperature (2988 K).

Thermal spike simulations are then performed to estimate the evolution of the lattice temperature upon the 27 MeV Fe impact. The motivation of these simulations is to evaluate the thermal spike response of the 900 keV I ($S_n$) defected S1300 sample in comparison to S600. Three sets of simulations are performed – (i) $S_n$ defected S1300, (ii) pristine S1300 (for comparison), and (iii) pristine S600. Since S600 is significantly less damaged than S1300 by the 900 keV I ions, we assume that the $S_n$ damage in S600 is negligible. Hence the simulations have been performed for pristine S600 only. For the pristine S1300 and S600 samples, the values of *K* and *g* are obtained by considering their respective grain sizes, in the manner described in our earlier work (Ref.[14]). The values of all the other relevant parameters are the same as taken in Refs.[14, 23]. To account for the increased scattering of the electrons and phonons from the defects in the $S_n$ defected S1300 system, the lattice thermal conductivity of this system is assumed to be an order of magnitude smaller than the pristine S1300 system, and the electron-phonon coupling constant for the system assumed to be 50% larger than the corresponding value for the pristine system[36-39]. Note that the value of these parameters *(K, g)* is affected by the level of disorder, i.e. the concentration of defects. It is however impossible to know the defect concentration (in the present case) after the incremental fluences/irradiations. Even if the defect concentration was known, there is no direct evidence of the exact relation between the values of these parameters *(g, K)* and the defect concentration[37-39]. We have thus assumed these values from literature (where the authors of these publications[36-39] have themselves taken these values on assumption). As such, the values of *K* and *g* for the defected S1300 system may not be very accurate and thus the values of the maximum thermal spike (lattice) temperature and thermal spike duration obtained from the thermal spike simulations may also not be very accurate. Nevertheless, these thermal spike simulations atleast give us an idea about the response of the defected S1300 system, in comparison to S600 system, to energy deposition by the Fe ions. The simulation results are shown in Figure 5, and the values of the maximum thermal spike temperature and thermal spike duration are listed in Table 3. As expected, the thermal spike in the case of the $S_n$ defected S1300 sample is notably more intense than in pristine S1300. Comparing S600 and the defected S1300 sample, although the maximum thermal spike temperature is higher for S600, the thermal spike duration is significantly (~ 200 ps) shorter. Since the $S_e$ induced radiation damage depends on both the (maximum) thermal spike temperature and its duration, the fact that the thermal spike duration is significantly longer for the already $S_n$ defected S1300 sample cannot be ignored (in analysing the response of the two systems to the energy deposition by the Fe ions). The thermal spike simulations thus suggest that, in addition to the S600 sample, the effect of $S_e$ is significant in case of the S1300 sample too under the simultaneous irradiation. This is in agreement with our model of the radiation damage as described above. Combined with the fact that S1300 is significantly damaged by $S_n$ as well (as opposed to S600), the simultaneous irradiations result in greater damage in S1300 as compared to S600.

|         | Maximum Temperature | Duration of thermal spike |
|---------|---------------------|---------------------------|
| S600    | ~ 8900 K            | ~ 100 ps                  |
| S1300   | ~ 3350 K            | ~ 10 ps                   |
| S1300 D | ~ 5000 K            | ~ 300 ps                  |

**Table 3:** Approximate values of maximum transient lattice temperature and duration of thermal spike at the center of the ion track.

## 5. Conclusions

In conclusion, we have investigated the effect of grain size on the radiation tolerance against 'simultaneous' nuclear and electronic energy loss. This was done by irradiating NC (S600) and bulk-like (S1300) YSZ samples with simultaneous 900 keV I ($S_n$) and 27 MeV Fe ($S_e$) ions. The irradiation induced damage is found to be lower in the NC sample.

The damage mechanism can be summarized as follows: (i) since there is essentially no spatial and time overlap between the damage events of the two ion beams, the simultaneous irradiation is actually a series of small sequential irradiations with incremental fluences; (ii) the NC S600 sample is essentially undamaged in comparison to the bulk-like S1300 against the incremental I ions (because of its smaller grain size); (iii) the subsequent impact by Fe ions (with incremental fluence) results in the formation of a damaged state for both S600 and S1300. The $S_e$ induced damage in the S600 sample is because of its small grain size. On the other hand, S1300 is also efficiently damaged by the $S_e$ (in-spite of its larger grain size) due to the existence of the defects, created earlier by $S_n$, that enhances the thermal spike; (iv) the irradiation damage in the S1300 sample thus consists of a superposition of the damage by $S_n$ and the damage by $S_e$ (which itself is a consequence of the pre-existing $S_n$ damage), while the S600 sample is essentially damaged only by $S_e$. The net result is that S1300 is more damaged than S600. Therefore, the better radiation tolerance of the NC sample is a consequence of its very nature (i.e. nano-crystallinity) itself.

Finally, it is worth emphasizing that the present results are important and provide the first steps towards the fundamental understanding of the interplay of grain size/GBs and combined $S_n$, $S_e$ in determining the radiation tolerance (against 'simultaneous' $S_n$ and $S_e$). The results also present a keen interest for the potential application of nano-crystalline materials in the nuclear industry where its nano-crystalline nature may result in the strong reduction of the damage production, thus allowing the conservation/prolongation of the physical integrity of the materials subjected to intense (simultaneous) irradiations. The influence of a couple of vital parameters, viz. irradiation temperature[14, 20, 23] and $S_e$ to $S_n$ ratio, however need to be investigated first for a complete fundamental understanding and before any (potential) applications.

## Acknowledgements

We sincerely acknowledge Dr. Rakesh Shukla of the Chemistry Division, Bhabha Atomic Research Centre and the JANNuS-Saclay facility staff for help during the sample preparation and irradiations respectively. We express gratitude towards Dr. Aurélien Debelle of CSNSM, Université Paris Saclay for the irradiation beam-time. We also express gratitude towards Prof. Pankaj Srivastava, Dept. of Physics, IIT Delhi for helpful discussions. Parswajit Kalita is thankful to MHRD, India (IIT Delhi) and IFCPAR/CEFIPRA (Raman-Charpak Fellowship) for financial assistantship. Santanu Ghosh acknowledges the financial support from BRNS, India (grant no. 37(3)/14/25/2017-BRNS/ 37222). XRD (Dept. of Physics) and Raman spectroscopy (CRF) facilities of IIT Delhi are acknowledged.